\documentclass[useAMS,usenatbib]{mn2e}
\usepackage[dvips]{graphicx}
\usepackage[english]{babel}
\usepackage{amsmath}
\usepackage{amssymb}
\usepackage{textcomp}
\usepackage{natbib}

\title[r-process enrichment by NS mergers]{Galactic r-process enrichment by neutron star mergers in cosmological simulations of a Milky Way-mass galaxy}
\author[F. van de Voort et al.]{Freeke~van~de~Voort$^{1,2}$\thanks{E-mail: freeke@berkeley.edu}, 
Eliot~Quataert$^{1}$, 
Philip~F.~Hopkins$^{3}$, 
Du\v{s}an~Kere\v{s}$^{4}$  
\newauthor
and Claude-Andr\'e~Faucher-Gigu\`ere$^{5}$ \\
$^{1}$Department of Astronomy and Theoretical Astrophysics Center, University of California, Berkeley, CA 94720-3411, USA \\
$^{2}$Academia Sinica Institute of Astronomy and Astrophysics, P.O. Box 23-141, Taipei 10617, Taiwan \\
$^{3}$TAPIR, Mailcode 350-17, California Institute of Technology, Pasadena, CA 91125, USA \\
$^{4}$Department of Physics, Center for Astrophysics and Space Science, University of California at San Diego, 9500 Gilman Drive, \\
\ \ La Jolla, CA 92093 \\
$^{5}$Department of Physics and Astronomy and CIERA, Northwestern University, 2145 Sheridan Road, Evanston, IL 60208, USA
}

\begin{document}
\newcommand\rp{\operatorname{r-process}}

\date{Accepted November 12, 2014; Received September 29, 2014; in original form July 25, 2014}

\pagerange{\pageref{firstpage}--\pageref{lastpage}} \pubyear{2014}

\maketitle

\label{firstpage}

\begin{abstract}

We quantify the stellar abundances of neutron-rich r-process nuclei in cosmological zoom-in simulations of a Milky Way-mass galaxy from the Feedback In Realistic Environments project. The galaxy is enriched with r-process elements by binary neutron star (NS) mergers and with iron and other metals by supernovae. These calculations include key hydrodynamic mixing processes not present in standard semi-analytic chemical evolution models, such as galactic winds and hydrodynamic flows associated with structure formation. We explore a range of models for the rate and delay time of NS mergers, intended to roughly bracket the wide range of models consistent with current observational constraints. We show that NS mergers can produce [r-process/Fe] abundance ratios and scatter that appear reasonably consistent with observational constraints. At low metallicity, $\mathrm{[Fe/H]}\lesssim-2$, we predict there is a wide range of stellar r-process abundance ratios, with both supersolar and subsolar abundances. Low-metallicity stars or stars that are outliers in their r-process abundance ratios are, on average, formed at high redshift and located at large galactocentric radius. Because NS mergers are rare, our results are not fully converged with respect to resolution, particularly at low metallicity. However, the uncertain rate and delay time distribution of NS mergers introduces an uncertainty in the r-process abundances comparable to that due to finite numerical resolution. Overall, our results are consistent with NS mergers being the source of most of the r-process nuclei in the Universe.

\end{abstract}

\begin{keywords}
nuclear reactions, nucleosynthesis, abundances -- stars: neutron -- stars: abundances -- Galaxy: abundances -- Galaxy: evolution -- methods: numerical
\end{keywords}

\section{Introduction}

The astrophysical site that produces elements heavier than iron by rapid neutron capture reactions (r-process) remains poorly understood. This is particularly true for the third peak in the abundance of r-process elements near atomic mass $\approx195$. The need for a high flux of free neutrons for rapid neutron capture reactions requires an astrophysical source with an abundance of free neutrons  \citep{BBFH}. Given this, the two astrophysical sites most commonly considered for the origin of r-process elements are core-collapse supernovae and binary neutron star (NS) (or neutron star--black hole) mergers.   

The neutrino-driven wind produced after a successful core-collapse supernova explosion was originally viewed as a promising site for the production of r-process nuclei \citep[e.g.][]{Meyer1992}.  However, increasingly sophisticated treatments of the dynamics and neutrino transport in neutrino-driven winds have systematically failed to produce the thermodynamic conditions necessary for  producing third-peak r-process nuclides \citep[e.g.][]{Qian1996,Thompson2001}.  This is even more true of many modern core-collapse simulations, which find  that the neutrino driven wind has a proton-rich rather than neutron-rich composition \citep[e.g.][]{Hudepohl2010}. Despite this general failure of core-collapse supernova models to produce r-process nuclei, there are sufficient remaining uncertainties that this conclusion is not yet definitive. In addition to uncertainties in neutrino physics and the core-collapse explosion mechanism, strong magnetic fields and rapid rotation change the thermodynamic conditions in neutrino-driven winds, making the conditions more favourable for the production of r-process nuclei \citep[e.g.][]{Metzger2007}.

In contrast to core-collapse supernovae, NS mergers can robustly produce third peak r-process nuclei. In particular, the decompression of neutron-rich matter from nuclear densities produces rapid neutron capture nucleosynthesis and a distribution of elements that is reasonably consistent with observations of the Sun and other stars \citep[e.g.][]{Lattimer1977,Freiburghaus1999}.    Unbound, neutron-rich matter is produced dynamically during NS mergers, in e.g. tidal tails, and in winds from the rotationally-supported accretion disk left after the merger \citep[e.g.][]{Lattimer1974,Metzger2009}.   

NS mergers are very likely to be the first source of gravitational waves directly detected by the ground-based interferometers Advanced LIGO and VIRGO \citep{Abbott2009, Accadia2012}. In this context, there is renewed motivation for assessing NS mergers as the source of r-process nuclei.  In particular, the decay of heavy elements produced during r-process nucleosynthesis powers isotropic electromagnetic emission that is predicted to be one of the most promising sources of electromagnetic emission coincident with a gravitational-wave detection \citep[e.g.][]{Metzger2010,Barnes2013}.  Constraints on r-process production during NS mergers thus directly constrains the properties of electromagnetic sources powered by r-process nucleosynthesis and, at least in principle, the rate of NS mergers themselves.

Given that NS mergers, unlike supernovae, robustly produce third peak r-process nuclei, a natural hypothesis is that the majority of the r-process elements in nature are produced by NS mergers. The observed abundances of r-process elements in galactic stars provide some of the most direct constraints on this hypothesis. Such observations show that even very metal poor stars have a solar r-process abundance pattern and that there is only modest scatter in the r-process to iron abundances of stars \citep[e.g.][]{Sneden2008}. In simple models of galactic chemical enrichment, these observations are difficult to reproduce \citep[e.g.][]{Qian2000, Argast2004, Matteucci2014}. This is because NS mergers are relatively rare compared to supernovae, which increases the scatter in the abundance of r-process elements and decreases the likelihood that early generations of stars (i.e.\ low-metallicity stars) are enriched by r-process nucleosynthesis.

However, existing quantitative models of galactic r-process enrichment rely on simplified chemical evolution calculations that do not capture the full complexity of how galaxies form. In particular, stellar feedback during galaxy formation drives powerful winds which redistribute heavy elements far from their birth locations \citep{Veilleux2005}. This mixing is not present in idealized chemical evolution calculations, which thus likely overestimate the star-to-star scatter in heavy-element abundances. 

Motivated by these potential deficiencies of existing models of galactic r-process enrichment, this paper studies the abundances of r-process elements using state-of-the-art galaxy formation simulations, which include galactic winds and hydrodynamic flows associated with structure formation (thus alleviating some of the problems that plague simple semi-analytical chemical evolution models). We explore a large range of parameters values in a simple model for r-process enrichment due to NS mergers, finding stellar abundance ratios in broad agreement with observations.  

In Section~\ref{sec:sim} we describe the simulation we used; the details of the different r-process element enrichment models can be found in Section~\ref{sec:models}. In Section~\ref{sec:results} we show the abundances for these models and how they correlate with stellar age and galactocentric distance. We discuss and summarize our results and their convergence in Section~\ref{sec:concl}.

\section{Method} \label{sec:sim}

We use a newly developed version of TreeSPH (P-SPH) which adopts the Lagrangian `pressure-entropy' formulation of the smoothed particle hydrodynamics (SPH) equations \citep{Hopkins2013PSPH}. The gravity solver is a heavily modified version of \textsc{gadget}-2 \citep[last described in][]{Springel2005}. P-SPH also includes substantial improvements in the artificial viscosity, entropy diffusion, adaptive timestepping, smoothing kernel, and gravitational softening algorithm.

This work is part of the Feedback in Realistic Environments (FIRE) project\footnote{http://fire.northwestern.edu/} \citep{Hopkins2014FIRE}, which consists of several cosmological `zoom-in' simulations. Here we make use of galaxy \textbf{m12i}, a galaxy with mass similar to the Milky Way at $z=0$. Its initial conditions are taken from the AGORA project \citep{Kim2014}. The simulation is described in \citet{Hopkins2014FIRE} and references therein and we will only summarize its main properties here.

This cosmological simulation assumes a $\Lambda$CDM cosmology with parameters consistent with the 9-yr Wilkinson Microwave Anisotropy Probe (WMAP) results \citep{Hinshaw2013}, $\Omega_\mathrm{m}= 1-\Omega_\Lambda = 0.272$, $\Omega_\mathrm{b}= 0.0455$, $h = 0.702$, $\sigma_8 = 0.807$ and $n = 0.961$. 

\begin{table*}
\begin{center}
\caption{\label{tab:sims} \small Simulation parameters: simulation resolution identifier, initial mass of gas particles ($m_\mathrm{baryon}$), mass of dark matter particles ($m_\mathrm{DM}$), minimum baryonic force softening ($\epsilon_\mathrm{baryon}$), minimum dark matter force softening ($\epsilon_\mathrm{DM}$), final simulation redshift ($z_\mathrm{final}$), number of star particles within 50~proper~kpc of the galaxy centre at $z_\mathrm{final}$ ($N_\mathrm{star}$).}
\begin{tabular}[t]{llllllr}
\hline
\hline \\[-3mm]
resolution & $m_\mathrm{baryon}$ (M$_\odot$) & $m_\mathrm{DM}$ (M$_\odot$) & $\epsilon_\mathrm{baryon}$ ($h^{-1}$pc) & $\epsilon_\mathrm{DM}$ ($h^{-1}$pc) & $z_\mathrm{final}$ & $N_\mathrm{star}$ \\
\hline \\[-4mm]                                                                                                                                       
low & $4.5\times10^5$ & $2.3\times10^6$ & 30 & 250 & 0 & 117,901 \\
fiducial & $5.7\times10^4$ & $2.8\times10^5$ & 14 & 100 & 0 & 611,995 \\                                
%fiducial & $5.7\times10^4$ & $2.8\times10^5$ & 2.4 & 43,255 \\                                
high & $7.1\times10^3$ & $3.5\times10^4$ & 14 & 100 & 2.4 & 410,995 \\ 
\hline                                                                                                                                                
\end{tabular}
\end{center}
\end{table*}   

Because of the expansion of the Universe and the collapse of structure, the volume of the zoom-in region depends on redshift, but is always much larger than the virial radius. It is about a Mpc in radius at $z=0$. The (initial) particle masses for baryons and dark matter are $5.7\times10^4$~M$_\odot$ and $2.8\times10^5$~M$_\odot$, respectively, for our fiducial resolution. We also consider simulations with eight (two) times lower and higher mass (spatial) resolution. Particle masses and minimum physical softening lengths for all the simulations used in this work are listed in Table~\ref{tab:sims}, as well as the final redshift of each simulation and the number of star particles included in our analysis, i.e.\ all star particles within 50~kpc of the centre of the galaxy. The minimum physical baryonic force softening length for our fiducial resolution is 14~$h^{-1}$pc. We adopt a quintic spline kernel with an adaptive size, keeping the mass within the kernel approximately equal. The average number of neighbours in the smoothing kernel is 62. 

Star formation takes place in molecular, self-gravitating gas above a hydrogen number density of $n_\mathrm{H}^\star>10$~cm$^{-3}$, where the molecular fraction is calculated following \citet{Krumholz2011} and the self-gravitating criterion following \citet{Hopkins2013SelfGrav}. Stars are formed at the rate $\dot\rho_\star=\rho_\mathrm{molecular}/t_\mathrm{ff}$, where $t_\mathrm{ff}$ is the free-fall time. The star formation parameters are kept fixed when we vary the resolution. Star particles inherit their metal abundances from their progenitor gas particle.

We assume an initial stellar mass function (IMF) from \citet{Kroupa2002}. Radiative cooling and heating are computed in the presence of the cosmic microwave background (CMB) radiation and the ultraviolet (UV)/X-ray background from \citet{Faucher2009}. Self-shielding is accounted for with a local Sobolev/Jeans length approximation. A temperature floor of 10~K or the CMB temperature is imposed. 

The primordial abundances are $X = 0.76$ and $Y = 0.24$, where $X$ and $Y$ are the mass fractions of hydrogen and helium, respectively. The simulation has a metallicity floor at $Z=10^{-4}$~Z$_\odot$, because yields are very uncertain at lower metallicities. The abundances of 11 elements (H, He, C, N, O, Ne, Mg, Si, S, Ca and Fe) produced by massive and intermediate-mass stars (through Type Ia supernovae, Type II supernovae, and stellar winds) are computed following \citet{Iwamoto1999}, \citet{Woosley1995}, and \citet{Izzard2004}. There is \emph{no} sub-resolution metal diffusion in our simulation.
Supernovae and stellar winds are modelled by having a fraction of the mass ejected by a star particle transferred to its neighbouring gas particle $j$ as follows:
\begin{equation}
f_j = \dfrac{\frac{m_j}{\rho_j} W(r_j,h_\mathrm{sml})}{\Sigma_i \frac{m_i}{\rho_i} W(r_i,h_\mathrm{sml})},
\end{equation}
where $h_\mathrm{sml}$ is the smoothing length of the star particle (determined in the same manner as for gas particles), $r_i$ is the distance from the star particle to neighbour $i$, $W$ is the quintic SPH kernel, and the summation is over all SPH neighbours of the star particle, i.e.\ 62 on average. 

The FIRE simulations include an implementation of stellar feedback by supernovae, radiation pressure, stellar winds, and photo-ionization and photo-electric heating (see \citealt{Hopkins2014FIRE} and references therein for details). For the purposes of the present paper, we emphasize that these simulations produce galaxies with stellar masses reasonably consistent with observations over a wide range of dark matter halo masses. This is a consequence of galactic winds driven by stellar feedback. The same winds also redistribute the heavy elements created in stars, both into the surrounding circum-galactic medium (CGM) and elsewhere in the galactic disk via a `galactic fountain.' This feedback-induced mixing is important for the chemical evolution of the galaxy.

The galaxy in our fiducial simulation is an isolated, star-forming, bulge-dominated galaxy. It builds up a disc component at low redshift ($z<0.5$), which kinematically comprises of about a third of the stars in the galaxy.  Since these stars are young, however, their contribution to the total stellar light (or light-weighted kinematics) is larger. Measured within the central 20~kpc at $z=0$, the star formation rate is about 3~M$_\odot$yr$^{-1}$, the stellar mass is $3\times10^{10}$~M$_\odot$, the total gas mass is $9\times10^{9}$~M$_\odot$, and the mass of the star-forming gas is $9\times10^{8}$~M$_\odot$. We have repeated our analysis for more disc-dominated as well as for more bulge-dominated galaxies within the FIRE sample \citep{Hopkins2014FIRE} and found our results for the r-process abundances to be very similar.

\subsection{r-process models} \label{sec:models}

R-process elements are implemented as tracers and have no effect on the simulation. As a result, we are able to implement multiple models of r-process enrichment in the same hydrodynamic simulation, saving considerable computational time.

We assume that NS mergers are responsible for all of the r-process enrichment. However, the rates of NS mergers are quite uncertain so we consider a variety of models to bracket this uncertainty. Using the known population of binary neutron stars in the Milky Way, \citet{Abadie2010} estimate that the merger rate is $\sim 10^{-4}$ yr$^{-1}$, but that the rate could range from $10^{-6}-10^{-3}$ yr$^{-1}$ under extreme assumptions.   For the purposes of understanding the enrichment of low-metallicity stars, it is also important to know the delay time distribution that characterizes the merger rate as a function of time after a burst of star formation.   Population synthesis calculations suggest a rate $\propto t^{-1}$ once lower mass core collapse supernovae have exploded, i.e., after a delay of $\sim 10^7$ yrs (e.g., \citealt{Belczynski2006}).   Given these constraints, we take the
rate of NS mergers ($R_\mathrm{NS}$) to be
\begin{equation} \label{eqn:rate}
R_\mathrm{NS} = \frac{A M_\mathrm{star}}{t} \ \mathrm{for} \ t > t_\mathrm{min}
\end{equation}
and $R_\mathrm{NS}=0$ otherwise, where $A$ is the number of NS mergers per unit of stellar mass, $M_\mathrm{star}$ is the stellar mass under consideration, $t$ is time since the formation of the star particle, and $t_\mathrm{min}$ is the minimum time needed for a NS merger to take place. In the simulations the rate of NS mergers is implemented stochastically, since we do not resolve individual stars. We take fiducial values of $A=10^{-5}$~M$_\odot^{-1}$ and $t_\mathrm{min}=3\times10^7$~yr. The stellar mass of our Milky Way model at $z = 0$ is about $3\times10^{10}$~M$_\odot$, with half of the stars having formed by $z\approx1$. Thus our fiducial choice of $A$ corresponds to a present day merger rate of $\sim10^{-4}$~yr$^{-1}$, similar to that suggested by observations. 
Given the significant uncertainties in the NS merger rate, however, we explore a range of models with $t_\mathrm{min}$ ranging from $3\times10^6$ to $3\times10^8$~yr and $A$ ranging from $3\times10^{-6}$ to $3\times10^{-5}$~M$_\odot^{-1}$ (see Table~\ref{tab:models}).  Although we focus on model 0, with $A=10^{-5}$~M$_\odot^{-1}$ and $t_\mathrm{min}=3\times10^7$~yr, we stress that all of the models that we consider are plausible and so provide an estimate of the systematic uncertainty in the chemical evolution of r-process elements implied by the uncertainty in the NS merger rate and delay time distribution.

Our calculations do not include NS kicks, though it would clearly be of interest to do so in future work. The median offset of short-duration gamma-ray bursts is modest, about 4.5~kpc, though $\sim25$ per cent of events have offsets greater than 10~kpc \citep{Fong2013}. We discuss this uncertainty further in Section~\ref{sec:concl}.

The amount of r-process elements ejected per NS merger is uncertain.  However, an ejecta mass $\sim10^{-2} M_\odot$ per merger together with a merger rate $\sim 10^{-4}$ yr$^{-1}$ in the Milky Way is sufficient to produce the total mass of r-process nuclei in the Galaxy, which is $\sim10^4 M_\odot$ assuming all stars have roughly solar abundances. 
Since in our simulation the r-process elements act only as tracer particles, we choose to simply normalize our six models to solar metallicity, i.e. to set [r-process/Fe]=0 when [Fe/H]=0. This should be a reasonable approximation unless there is a significant systematic variation in the r-process yields per NS merger with redshift or stellar population age.

Another important parameter characterizing r-process enrichment is the mass of the ambient interstellar medium (ISM) that is initially enriched with r-process elements. Treating the ejecta from NS mergers as an expanding remnant analogous to a supernova remnant, the final momentum of the remnant during the momentum conserving phase is \citep{Cioffi1988}
\begin{equation}\label{eqn:momentum}
M v|_\mathrm{final} \approx 6 \times 10^4 E_{50}^{13/14} n_0^{-1/7}~\mathrm{M_\odot\,km\,s^{-1}},
\end{equation}
where $n_0$ is the ambient density in cm$^{-3}$ and we have scaled the initial energy of the remnant, $E_{50}$, to correspond to an ejecta mass of $0.01$~M$_\odot$ moving at $v = 0.1$~c, i.e. $10^{50}$~ergs \citep{Piran2013}. Equation~\ref{eqn:momentum} is for solar metallicity, but the final momentum is only about twice as large for a primordial composition. The total swept-up mass when the neutron-rich ejecta comes into approximate pressure equilibrium with the ambient medium is set by evaluating Equation~\ref{eqn:momentum} given a final velocity comparable to the sound speed or turbulent velocity of the ambient medium. This suggests that the neutron-rich ejecta is initially incorporated into $M_{\rm swept} \sim 10^{3.5-4}$~M$_\odot$ of the ambient ISM, depending on the exact ambient density and turbulent velocity. 

In simple chemical evolution calculations, the neutron-rich ejecta are typically taken to mix with \emph{only} a mass $M_{\rm swept}$ \citep[e.g.][]{Qian2000, Argast2004}. It is unlikely, however, that this is correct, given the vigorous turbulent mixing induced by stellar feedback, galactic winds, galactic fountains, galaxy mergers, instabilities in the differentially rotating galactic disc, etc. (as also argued by \citealt{Piran2014}).
To address this, one would ideally like to have a hydrodynamic simulation with a mass resolution comparable to $M_{\rm swept}$ since the simulation would (at least in principle) self-consistently resolve additional mixing produced by galaxy-scale turbulence and winds. This is somewhat prohibitive, however. Our fiducial simulation has a baryonic particle mass of $5.7 \times 10^4 M_\odot$ while our highest resolution simulation discussed in Section~\ref{sec:res} has a baryonic particle mass of $7 \times 10^3$~M$_\odot$. However, the mass in the kernel -- which is the mass to which the metal enrichment is applied -- is roughly 62 times larger. In Section~\ref{sec:concl} we show and discuss the dependence of our results on resolution. It is important to bear in mind that in all of our simulations the r-process enrichment initially occurs in a mass somewhat larger than the initial mass swept-up by a NS merger remnant. As we show below, the uncertainties in r-process enrichment associated with uncertainties in the NS merger rate are comparable to those due to the finite numerical resolution of our simulations.

\begin{table}
\begin{center}                                                                                                                                        
\caption{\label{tab:models} \small Parameters of r-process enrichment models: model number, number of NS mergers per $M_\odot$ of stars (see Eq.~\ref{eqn:rate}), minimum time needed for the first NS merger (see Eq.~\ref{eqn:rate}), NS merger rate at $z=0$. For each model we vary only one parameter, indicated in bold.}         
\begin{tabular}[t]{crrc}
\hline
\hline \\[-3mm]                                                                                                                                       
model & $A$ [M$_\odot^{-1}$] & $t_\mathrm{min}$ [yr] & $R_\mathrm{NS}(z=0)$ [yr$^{-1}$]\\
\hline \\[-4mm]                                                                                                                                       
0 & $10^{-5}$ & $3\times10^7$ & $1.5\times10^{-4}$\\
1 & $10^{-5}$ & $\mathbf{3\times10^6}$ & $2.1\times10^{-4}$\\                                                          
2 & $10^{-5}$ & $\mathbf{10^7}$ & $ 1.8\times10^{-4}$\\
3 & $10^{-5}$ & $\mathbf{10^8}$ & $1.1\times10^{-4}$ \\
4 & $\mathbf{3\times10^{-6}}$ & $3\times10^7$ & $4.4\times10^{-5}$\\
5 & $\mathbf{3\times10^{-5}}$ & $3\times10^7$ & $4.4\times10^{-4}$\\
\hline                                                                                                                                                
\end{tabular}                                                                                                                                         
\end{center}                                                                                                                                          
\end{table}      

\section{Results} \label{sec:results}

Throughout the paper we express abundance ratios of a star compared to those of the Sun as
\begin{equation}
[A/B]=\mathrm{log}_{10}\left(\frac{N_A}{N_B}\right)_\mathrm{star}-\mathrm{log}_{10}\left(\frac{N_A}{N_B}\right)_\odot,
\end{equation}
where $A$ and $B$ are different elements, $N_{\mathrm{A}}$ and $N_{\mathrm{B}}$ are number densities. Given that our simulations use a metallicity floor at $Z=10^{-4}$~Z$_\odot$ for all metal species, except tracer species, [Fe/H]$\ge-4$ and [Mg/H]$\ge-4$. We therefore have to be careful interpreting our results at low metallicity. Results are shown for [Fe/H]$\ge-3.5$, a factor three above the metallicity floor, where our results are not strongly affected by the choice of metallicity floor. The initial conditions, however, do not include a metallicity floor for r-process elements, so particles can have $\mathrm{[\rp/Fe]}=-\infty$. We take those particles into account as well when we calculate the median and percentiles below.

\begin{figure}
\center
\includegraphics[scale=.46]{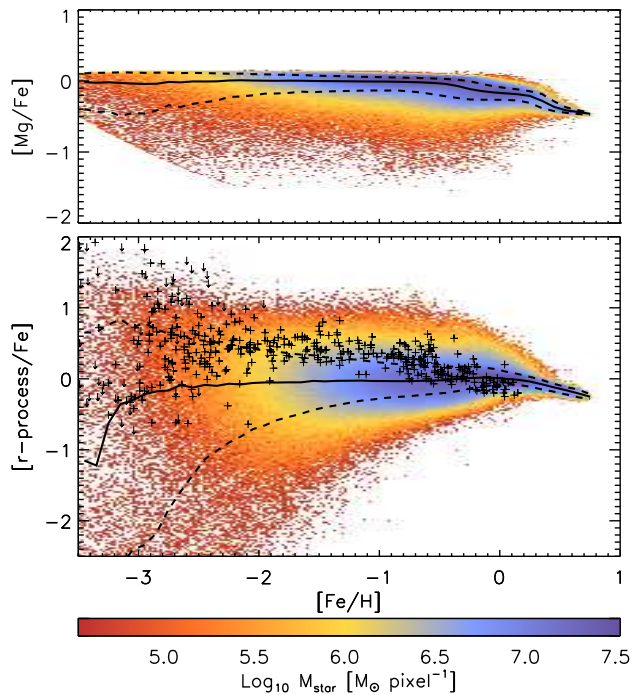}
\caption {\label{fig:rpFe} [Mg/Fe] (top panel) and [r-process/Fe] (bottom panel) as a function of [Fe/H] at $z=0$ for our fiducial resolution simulation. The colour coding indicates the logarithm of the stellar mass per pixel and both images are 200 by 200 pixels. Black curves show the median (solid curves) and 16th and 84th percentiles (dashed curves). [r-process/Fe] has been normalized so that its median is zero at $\mathrm{[Fe/H]}=0$.   [Mg/Fe] is not normalized but is set by the supernova yields.   Black plusses and downward arrows are observed [Eu/Fe] detections and upper limits, respectively \citep{Suda2008}. The median [Mg/Fe] and [r-process/Fe] are both fairly constant with metallicity, although they decrease slightly at $\mathrm{[Fe/H]}>0$. [r-process/Fe] decreases strongly at $\mathrm{[Fe/H]}<-3$. The scatter in [Mg/Fe] is small at all metallicities, but the scatter in [r-process/Fe] increases significantly towards lower metallicity.}
\end{figure}

Figure~\ref{fig:rpFe} shows the $z=0$ abundance ratios of Mg (top panel) and r-process elements (bottom panel) to Fe as a function of metallicity for our fiducial resolution simulation. We include all star particles at radii $R<50$~kpc from the centre of the galaxy, but our results do not depend on this choice. Black curves show the median (solid curves) and 16th and 84th percentiles (dashed curves). As described in Section~\ref{sec:models}, [r-process/Fe] has been normalized so that its median is zero at $\mathrm{[Fe/H]}=0$. [Mg/Fe] is computed directly from the heavy element yields used in the simulation and is not normalized. The bottom panel of Figure~\ref{fig:rpFe} also shows observational data points for [Eu/Fe], where plusses show detections and downward arrows show upper limits \citep{Suda2008}.\footnote{We exclude carbon-enhanced metal-poor stars, since their origin is not well understood.} Eu is an ideal tracer of r-process enrichment, because it is almost solely produced in r-process events \citep{Burris2000}. It is not possible to compare our models and the observations in detail, because of unknown selection effects in the heterogeneous observational sample. Global trends are likely robust.

Mg is the $\alpha$ element most easily observed and therefore most useful for comparison with observations. It is thought to be primarily produced in Type II supernovae. Consistent with observations of the Milky Way \citep[e.g.][]{Cayrel2004, Arnone2005}, we find very little scatter in [Mg/Fe] for both metal-poor and metal-rich stars, little variation in the median at $\mathrm{[Fe/H]}<-1$ and a small decrease at high metallicity, due to the increasing importance of Type Ia supernovae, which mainly produce iron-peak elements. The median [Mg/Fe] in our simulation is, however, lower by about 0.3~dex as compared to observations. This discrepancy is well-known and is due to the \citet{Woosley1995} yields for Mg being too low by a factor of a few \citep[e.g.][]{Timmes1995, Thomas1998, Portinari1998, Lia2002}. Because Mg is not a dominant coolant in our simulation, this discrepancy in Mg abundance has no effect on the evolution of the galaxy and could be straightforwardly fixed by adopting better yields in future simulations or by increasing the Mg abundance in post-processing.  We choose to leave the Mg abundances as is in Figure~\ref{fig:rpFe} to accurately reflect the actual results in our simulation.  Moreover, the trends in Mg abundance with metallicity and the scatter in Figure~\ref{fig:rpFe} are robust, since Mg is almost purely produced in Type II supernovae.  In future work, we plan to compare the abundance ratios of the various metal species, the mass-metallicity relation, and metallicity gradients in our different galaxies to observations.  We include the Mg results in the present paper primarily to highlight the different scatter relative to the r-process elements.

As is the case for Mg, the decrease of [r-process/Fe] at $\mathrm{[Fe/H]}>0$ in Figure~\ref{fig:rpFe} is due to Fe production by Type Ia supernovae, which become relatively more important at late times. This is also consistent with observations, although there the decrease sets in at somewhat lower metallicities. This could be due to selection biases in the observations or to the absense of certain mixing processes in our simulation. See Section \ref{sec:concl} for more discussion on the latter possibility. The median [r-process/Fe] is approximately flat at $\mathrm{[Fe/H]}<0$ in the simulation at our fiducial resolution, whereas the data show a small increase to $\mathrm{[\rp/Fe]}\approx0.5$. In both the observations and simulation the scatter increases substantially towards low metallicity. Note that many of the low-metallicity observations are upper limits (shown as downward arrows). Overall, our simulation is in reasonable agreement with observations, especially taking into account the uncertainties in the metal yields as well as the unknown selection effects in the data. 

\begin{figure}
\center
\includegraphics[scale=.5]{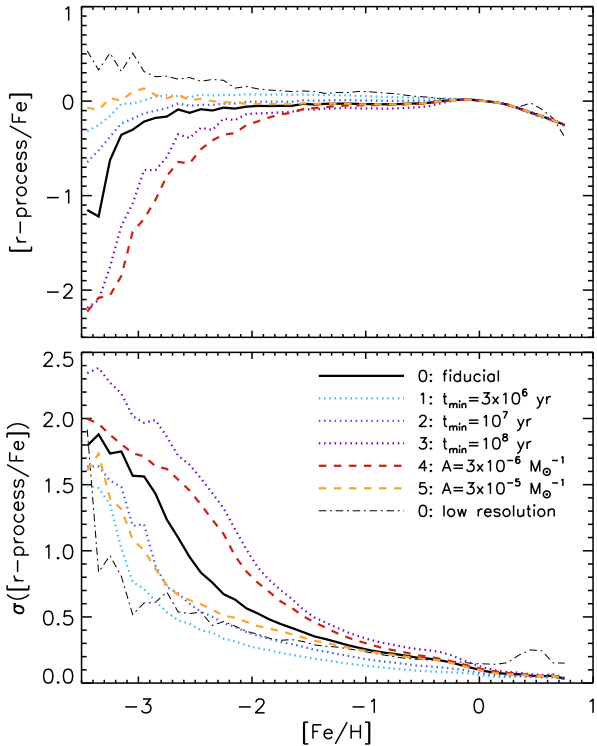}
\caption {\label{fig:models} Median [r-process/Fe] (top panel) and scatter ($\sigma$) in [r-process/Fe] (bottom panel) as a function of [Fe/H] at $z=0$ for the six different models listed in Table~\ref{tab:models} and indicated in the legend. The black, solid curve is for model 0, also shown in Figure~\ref{fig:rpFe}. The thin, black, dot-dashed curve is model 0 for a low resolution simulation, described in Section~\ref{sec:res}. Dotted curves show variations of $t_\mathrm{min}$. Dashed curves show variations of $A$. As $t_\mathrm{min}$ increases or $A$ decreases, [r-process/Fe] decreases and the scatter increases, especially at lower metallicities.}
\end{figure}

As mentioned before, the NS merger rate is highly uncertain and we therefore explore different NS merger models. Figure~\ref{fig:models} shows the median [r-process/Fe] (top panel) and scatter in [r-process/Fe] (bottom panel) as a function of [Fe/H] for the six different models listed in Table~\ref{tab:models}. The solid, black curve is identical to the one in Figure~\ref{fig:rpFe}. The scatter is calculated by subtracting the 16th percentile from the 84th percentile (dashed curves in Figure~\ref{fig:rpFe}) and dividing by two. Dotted curves show variations of $t_\mathrm{min}$. Dashed curves show variations of $A$, as indicated in the legend. We reiterate that given the uncertainties in the NS merger rate, all of these models are plausible.

The different models span a range of median [r-process/Fe] values of only $0.4$~dex at $\mathrm{[Fe/H]}=-2$, but differ significantly at lower metallicity. With higher $t_\mathrm{min}$, i.e.\ a larger delay time before the first NS merger takes place, or lower $A$, i.e.\ fewer NS mergers per unit of stellar mass, the median [r-process/Fe] decreases at low metallicity, the drop in r-process element abundances moves to higher metallicity, and the scatter in [r-process/Fe] increases. This is due to NS mergers becoming more rare, which increases the stochasticity. Vice versa, when we decrease $t_\mathrm{min}$ or increase $A$, [r-process/Fe] increases, the low-metallicity drop moves to lower metallicities, and the scatter decreases. For all our models, the median and scatter are large enough that an appreciable fraction of low-metallicity stars has high r-process abundance ratios. The 84th percentile of the distribution has $[\rp/\mathrm{Fe}]>0$ for all models. 

\begin{figure}
\center
\includegraphics[scale=.5]{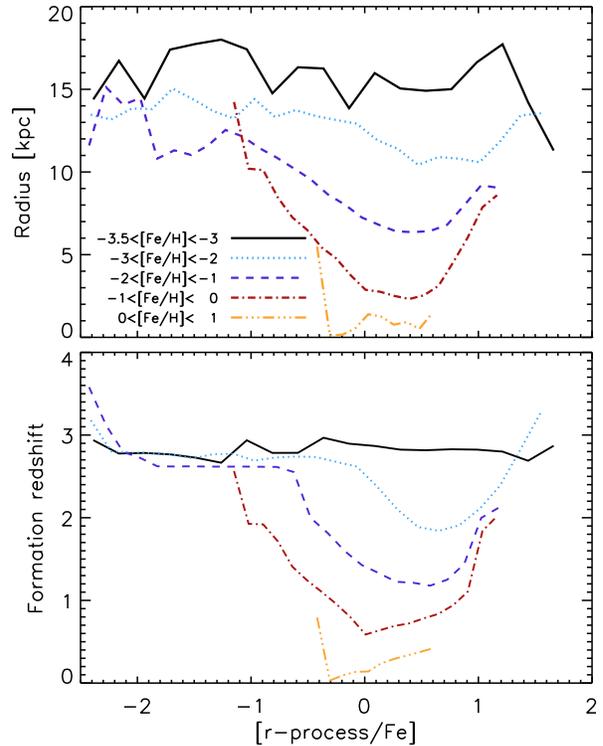}
\caption {\label{fig:zradius} Median galactocentric radius at $z=0$ (top panel) and median redshift at which the stars were formed (bottom panel) as a function of [r-process/Fe] for different [Fe/H] bins as indicated in the legend. As the metallicity increases, the distance from the galaxy centre and the stellar age decrease. At fixed metallicity, distance and age are lowest for stars with $\mathrm{[\rp/Fe]}\approx0-0.5$ and increase as the absolute value of [r-process/Fe] increases. This shows that the stars responsible for the large scatter in [r-process/Fe] are, on average, old and currently in the halo of the galaxy.}
\end{figure}
Lower-metallicity stars and stars with [r-process/Fe] more different from solar are, on average, older and located at larger distances from the centre of the galaxy. This is quantified in Figure~\ref{fig:zradius}, where the median galactocentric radius ($\langle R\rangle$, top panel) and the median formation redshift ($\langle z_\mathrm{form}\rangle$, bottom panel) of stars are shown as a function of [r-process/Fe]. The iron abundance increases for the different curves from top to bottom as indicated by the legend. The star particles have been selected in the same way as for Figs.~\ref{fig:rpFe} and~\ref{fig:models}, i.e.\ within 50~kpc of the centre of the galaxy. 

Stars with supersolar metallicities are found close to the centre and have been formed very recently, with $\langle R\rangle=0.7$~kpc and $\langle z_\mathrm{form}\rangle=0.1$. Radius and formation redshift increase with decreasing stellar metallicity. Stars with $-3.5<\mathrm{[Fe/H]}<-3$ have $\langle R\rangle=16$~kpc and $\langle z_\mathrm{form}\rangle=2.9$. This is quantitatively consistent with observations of the Milky Way, where halo stars are found to be older and more metal poor than disc stars. Note that most of our low-metallicity stars were not accreted, but formed in situ at high redshift and moved to larger radii through dynamical effects. 

At high metallicity, stars with $\mathrm{[\rp/Fe]}\approx0$ have the smallest $R$ and $z_\mathrm{form}$. This minimum moves to $\mathrm{[\rp/Fe]}\approx0.6$ for $-3<\mathrm{[Fe/H]}<-2$. Outliers in [r-process/Fe] are found at larger radii and are older, on average. For example, for $\mathrm{[Fe/H]}\approx-1$,  $\langle R\rangle=2$~kpc and $\langle z_\mathrm{form}\rangle=0.3$ at $\mathrm{[\rp/Fe]}\approx0$, but at $\mathrm{[\rp/Fe]}\approx-1$, $\langle R\rangle=8$~kpc and $\langle z_\mathrm{form}\rangle=2.1$. This indicates that the scatter is driven by inhomogeneous chemical evolution at high redshift. Stars formed at high redshift are moved to larger radii through dynamical effects. The dependence of $\langle R\rangle$ and $\langle z_\mathrm{form}\rangle$ on [r-process/Fe] disappears for stars with $\mathrm{[Fe/H]}<-3$.

\subsection{Resolution test} \label{sec:res}

As discussed in Section~\ref{sec:models}, the mass of the ISM initially enriched by r-process nuclei is one of the key parameters setting the chemical evolution of neutron-rich nuclei. Typically it is assumed that the r-process elements mix only with the initial swept-up mass in the NS merger remnant \citep[e.g.][]{Qian2000, Argast2004}. Given the many sources of turbulent mixing in galaxies, however, it is by no means clear that this physical picture is correct. It is, however, the case that due to limited resolution, the r-process nuclei in our simulations initially enrich an ISM mass larger than the ejecta mass swept up prior to the NS merger remnant reaching pressure equilibrium with its surroundings (see Eq.~\ref{eqn:momentum} and associated discussion). In particular, we reiterate that in our SPH formulation all gas particles inside the kernel of a given star particle are enriched, i.e.\ 62 on average, weighted according to their SPH weights. Therefore, the enriched mass, i.e.\ about $3.5\times10^6$~M$_\odot$, is approximately $62$ times larger than the particle mass for our fiducial resolution, although it is not enriched uniformly. To test the sensitivity of our results to the mass resolution, we ran a simulation with 8 (2) times \emph{lower} mass (spatial) resolution down to $z=0$ and repeated our analysis. We checked that the galoctocentric distances and formation redshifts of stars shown in Figure~\ref{fig:zradius} are similar to the ones at low resolution. The Mg abundances and scatter are also very similar in both simulations. We conclude that these properties are converged.

The lower resolution results for model 0 are shown as thin, dot-dashed curves in Figure~\ref{fig:models}. Comparing to the fiducial resolution, we see that our r-process abundances are not completely converged. At low resolution, [r-process/Fe] decreases slightly towards $\mathrm{[Fe/H]}=0$ and there is no drop at low metallicity.\footnote{We do see a drop at this resolution for model 4, at $-3.5<\mathrm{[Fe/H]}<-3$, i.e.\ at lower metallicity than for our fiducial resolution simulation.} The scatter is lower at $\mathrm{[Fe/H]}<-1$, but still appreciable. This is not surprising, because lower resolution simulations mix the r-process elements into a larger initial mass, i.e.\ the smoothing kernel, reducing the dispersion in [r-process/Fe]. We note, however, that the lower resolution results are similar to some of our physically plausible variants at the fiducial resolution, in particular the models with more NS mergers per M$_\odot$ of stars (model 5) or shorter delay times (models 1 and 2).

\begin{figure}
\center
\includegraphics[scale=.5]{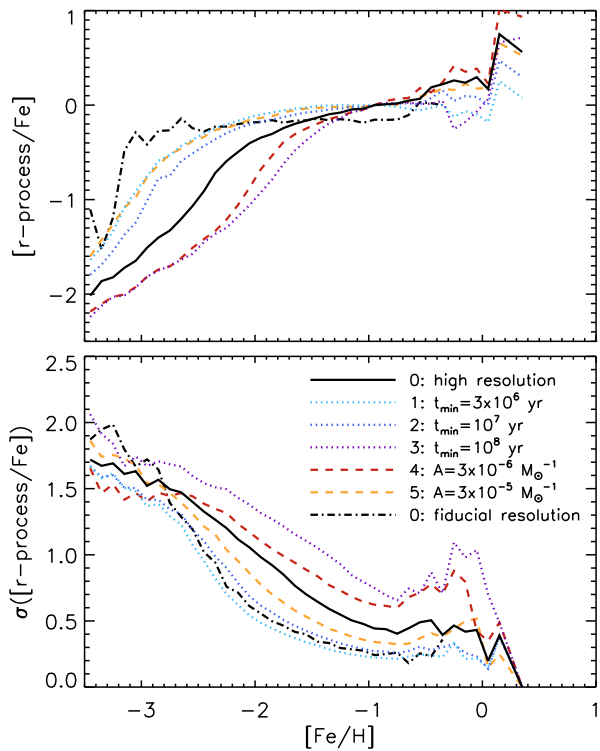}
\caption {\label{fig:models_z2p4} Same as Figure~\ref{fig:models}, but for a high resolution simulation at $z=2.4$. The black, dot-dashed curve is model 0 for our fiducial resolution at $z=2.4$. We normalized [r-process/Fe] at $\mathrm{[Fe/H]}=-1$ due to the low number of stars at higher metallicities.  At higher resolution, the drop in median [r-process/Fe] moves to higher metallicity and becomes less pronounced; the scatter in [r-process/Fe] also increases. Our different models for the NS merger rate show significant variations for the median [r-process/Fe] at low metallicity, as well as for scatter in [r-process/Fe] at $\mathrm{[Fe/H]}>-3$. Models 1, 2, and 5 at high resolution are reasonably similar to the fiducial resolution model 0.
}
\end{figure}

We also ran a high resolution simulation with 8 (2) times \emph{higher} mass (spatial) resolution, but only down to $z=2.4$, since it is more computationally expensive. Figure~\ref{fig:zradius} shows that the low-metallicity stars are formed at high redshift and thus the abundance ratios at the low-metallicity end are already in place at $z=2.4$. Figure~\ref{fig:models_z2p4} shows the median [r-process/Fe] and scatter for our high-resolution simulation at $z=2.4$. We normalized [r-process/Fe] at $\mathrm{[Fe/H]}=-1$ due to the low number of stars at higher metallicities. The lack of sufficiently many stars also causes the results to be somewhat noisy at high metallicity. The black, dot-dashed curve in Figure~\ref{fig:models_z2p4} is model 0 for our fiducial resolution at $z=2.4$. Comparing it to the solid line in Figure~\ref{fig:models} shows that there is indeed not much evolution in the r-process element abundances at low metallicity from $z = 2.4$ to $z = 0$.

Figure~\ref{fig:models_z2p4} shows that at higher resolution, the drop in median [r-process/Fe] moves to higher metallicity ($\mathrm{[Fe/H]}\approx-2$) and becomes less abrupt. The scatter increases at $\mathrm{[Fe/H]}>-3$.  However, the 84th percentile of the distribution remains supersolar at all metallicities. Moreover, models 1, 2, and 5 at high resolution are reasonably similar to the fiducial resolution model 0, with the median [r-process/Fe] only decreasing at $\mathrm{[Fe/H]}<-2.5$, perhaps more consistent with the observations shown in Figure~\ref{fig:rpFe}.

The resolution tests shown here demonstrate that simulated r-process element abundance ratios are sensitive to the resolution of the simulation at $\mathrm{[Fe/H]}\lesssim-2$ (for the resolutions we were able to explore). The qualitative change in our results with resolution is similar to what one would expect on physical grounds, namely that higher mass resolution generally leads to somewhat larger scatter and a lower median r-process abundance at low metallicity. Comparing to the observational data shown in Figure~\ref{fig:rpFe}, Figure~\ref{fig:models_z2p4} shows that our higher resolution simulations seem to match the observed median slope of [r-process/Fe] as a function of metallicity less well. This could be an indication of extra mixing processes these simulations are not capturing; see Section 4 for a discussion. Again, the observations are very heterogeneous with unknown selection effects, so a direct comparison is non-trivial.  This sensitivity to resolution is, however, specific to NS merger models and is not present in our [Mg/Fe] or [Fe/H] abundance ratios, which are produced by supernovae. The difference, of course, is that NS mergers are much rarer. Another key result of our resolution tests is that the uncertainty in the NS merger rate and the delay time distribution produces an uncertainty in the r-process enrichment that is comparable to the differences between our fiducial and high-resolution simulations.

\section{Discussion and Conclusions} \label{sec:concl}

We have quantified the abundance pattern of neutron-rich r-process nuclei in cosmological zoom-in simulations of a Milky Way-mass galaxy enriched by simplified implementations of binary neutron star (NS) mergers. We have compared the r-process nucleosynthesis to that of Mg and Fe, which are predominantly produced by Type II and Ia supernovae, respectively. We have explored a range of models for the rate and delay time of NS mergers, intended to roughly bracket the wide range of models consistent with the known binary NS population and population synthesis calculations of merger rates (see \S~\ref{sec:models}). The r-process elements are passive tracers that do not affect the simulation dynamics in any way. Our primary conclusions are:

\begin{itemize}
\item Neutron star (NS) mergers can produce [r-process/Fe] abundance ratios and scatter that are broadly consistent with observations for stars with $-2\lesssim \mathrm{[Fe/H]}\lesssim 0$.
\item The uncertain rate and delay time distribution of NS mergers results in a large uncertainty in the r-process abundances. Some of our assumed NS merger models resemble available observations better than others. 
\item The results for the r-process abundance ratios are not fully converged, particularly at low metallicity, i.e.\ at $\mathrm{[Fe/H]}\lesssim-2$.   This is physical and is due to the low rate of NS mergers and the difficulty resolving mixing on scales of the ISM of galaxies even in our highest resolution cosmological zoom-in simulations.
\item At low metallicity, the scatter in [r-process/Fe] is large enough to explain the r-process abundances of observed low-metallicity stars. However, we also predict a large population of stars with significantly subsolar [r-process/Fe], which have not yet been observed.   The existence of this population depends sensitively on the rate of turbulent mixing in the ISM.  It is very likely that our simulations do not capture all of the key mixing processes and so may overestimate the number of low-metallicity stars with  subsolar [r-process/Fe].
\item The low-metallicity stars in our simulations are, on average, formed at high redshift ($z=2-3$) and large galactocentric radius ($\langle R\rangle=10-20$~kpc).
\item The scatter in [r-process/Fe] at fixed metallicity is driven by old stars at large distances from the centre of the galaxy. 
\end{itemize}

The motivation for studying chemical enrichment with simulations of galaxy formation is that the simulations include a wide variety of mixing processes not present in phenomenological chemical evolution models. In particular, our galaxy formation simulation includes physically motivated (but still subgrid on scales smaller than giant molecular clouds) treatments of stellar feedback that lead to stellar masses of galaxies comparable to those observed over a wide range of dark matter halo masses \citep{Hopkins2014FIRE}. This consistency is a consequence of galactic winds efficiently redistributing gas (and with it heavy elements) from the scale of the galaxy to much larger radii in the halo.  Our simulations are also based on a formulation of SPH, called P-SPH, that resolves the historical problems of SPH in capturing hydrodynamic instabilities that are important for mixing, in particular the Kelvin-Helmholtz and Rayleigh-Taylor instabilities \citep{Hopkins2013PSPH}.

We suspect that the two most important mixing processes in our simulations for the purposes of stellar abundance patterns are galactic winds and subsequent re-accretion of previously enriched gas from the surrounding halo and galactic fountains redistributing mass throughout the galaxy. Instabilities within the differentially rotating galactic disc, galaxy mergers, and gas accretion can also contribute to the mixing of the ISM.
In addition, there is turbulent mixing within the galaxy itself, although this is not well-resolved in fully cosmological simulations like those we have carried out here.  As we discuss below, our results differ from previous analyses of r-process enrichment (e.g., \citealt{Argast2004}), suggesting that these additional mixing processes are indeed important.  Despite the enhanced mixing introduced by stellar feedback, there is an important sense in which our calculations \emph{underestimate} the mixing of heavy elements: metals in our simulation are stuck to gas particles and do not diffuse to neighbouring gas particles.

Material unbound during NS mergers is the only known astrophysical site with conditions that robustly produce r-process nuclei.  In particular, current models of nucleosynthesis in core-collapse supernovae fail to produce the requisite conditions, although the uncertainties remain large \citep[e.g.][]{Qian1996,Thompson2001, Hudepohl2010}. 
The primary objections to NS mergers as the source of r-process nuclides have been on nucleosynthetic grounds. First, observations of stars with a range of metallicities find that heavy r-process nuclei are also accompanied by elements with atomic masses of ~90-120 \citep{Sneden2008, Qian2007}. This does not occur in the very low electron fraction conditions typically considered in NS merger ejecta, which do not produce nuclei with atomic mass ~90-120. However, recent work has shown that accretion disc outflows produced from the disc left after a NS merger have higher electron fractions than the tidal tail unbound during the dynamical phase of the merger \citep[e.g.][]{Metzger2008, Just2014}. These two sources of nucleosynthesis during NS mergers can thus satisfy the observational requirements on the source of neutron-rich heavy elements.  
The second nucleosynthetic objection to NS mergers as the origin of the r-process elements is that the rarity of NS mergers would lead to too much scatter in the star-to-star r-process abundances and the absence of low-metallicity stars with solar r-process abundances (e.g., \citealt{Argast2004}).  Our calculations significantly alleviate this objection, although with some remaining caveats discussed below.

One of our primary conclusions is that NS mergers can produce [r-process/Fe] abundance ratios and scatter that appear reasonably consistent with observations for stars with $-2\lesssim \mathrm{[Fe/H]}\lesssim 0$ (see Fig.~\ref{fig:rpFe}). This suggests that NS mergers may indeed account for most of the r-process nuclei in the Universe. In addition, the scatter in [r-process/Fe] increases with decreasing [Fe/H] in our calculations, as is also observed. In many of our models, there is a decrease in the median stellar [r-process/Fe] ratio at $\mathrm{[Fe/H]}\lesssim-2$ (see Fig.~\ref{fig:models}).  Even so, there are still many stars with high abundance ratios of r-process elements at all subsolar metallicities we can probe. In particular, for all of our models the 84th percentile in the [r-process/Fe] distribution is supersolar at all metallicities. This is in contrast to the simplified chemical evolution models of \citet{Argast2004}, which find a very sharp drop in [r-process/Fe] at $\mathrm{[Fe/H]}\approx-2$ and no stars at $\mathrm{[Fe/H]}\lesssim-3$ that are enriched with r-process elements.

A more quantitative comparison to observations will require understanding the observational selection effects that enter into existing samples of r-process element abundance measurements. For example, we find that stars of metallicity $-2<\mathrm{[Fe/H]}<0$ show a clear trend in increasing galactocentric radius with decreasing [r-process/Fe] abundance (see Fig.~\ref{fig:zradius}), so that any kinematic selection effects for halo stars would bias the inferred median [r-process/Fe] ratios. In the future, quantifying these effects using a direct comparison between our simulations and observations is likely to be a fruitful way of testing the NS merger origin of r-process nuclei. A specific test of our models would be additional observations of low-metallicity stars ($\mathrm{[Fe/H]}\lesssim-2$).  All of our calculations show that there should be some low-metallicity stars with $[\rp/\mathrm{Fe}]<-1$ (see e.g.\ Fig.~\ref{fig:rpFe}). If such stars are in fact completely lacking observationally, this would rule out the models described in this paper and point to either a different origin for r-process elements in the lowest metallicity stars or to physical processes not included in our simulation, such as additional turbulent diffusion. Turbulent diffusion results in fewer low-metallicity stars and a reduction of the scatter in both [Mg/Fe] and [r-process/Fe]. 

The primary uncertainty in our conclusions is that our results for the stellar r-process abundance patterns are not fully converged, particularly at low metallicity, i.e.\ at $\mathrm{[Fe/H]}\lesssim-2$. This is not surprising and is a consequence of the rarity of NS mergers as a source of heavy elements. By contrast, the abundance ratios of Mg to Fe are numerically converged. At our highest resolution, the gas particle mass is $7 \times 10^3 M_\odot$ so that the total mass of gas (within the SPH kernel) that is enriched per `merger' is on average about $4 \times 10^5$~M$_\odot$.  This is significantly larger than the mass of the ISM swept-up by the ejecta created during a NS merger (see Sec.~\ref{sec:models}). It is not at all clear, however, that the latter is a reasonable model for the mass of the ISM that the r-process material is mixed into, given the many additional sources of mixing in the ISM \citep[e.g.][]{Yang2012} and the fact that mixing on scales smaller than the scale height of the galactic disk is not well resolved in our simulations.  Thus, it may in fact be the case that our simulations describe a reasonable `initial' mass of the ISM enriched by r-process nuclei.  Ultimately, however, it will be necessary to carry out even higher-resolution simulations, particularly for understanding the r-process abundances at low metallicity.  Moreover, either grid-based calculations or SPH simulations with explicit metal diffusion at the scale of the kernel would be useful for determining the extent to which small-scale mixing modifies the convergence of the abundances at low metallicity. In the absence of such mixing the results cannot converge at the lowest metallicity in an SPH simulation in which metals are locked into gas particles. In addition, high resolution simulations of individual neutron star merger remnants mixing into a turbulent ISM would also be  valuable. Larger turbulent velocities \emph{decrease} the mass of the ISM swept up prior to the NS merger remnant reaching pressure equilibrium with its surroundings (Eq.~\ref{eqn:momentum}) but \emph{increase} the resulting turbulent mixing of the r-process enriched material with the surrounding ISM.   

We find that the variations in [r-process/Fe] abundances introduced by considering a range of plausible NS merger rates and delay times is comparable to or larger than the changes introduced by numerical resolution.  Specifically, if the delay time for a significant fraction of NS mergers is $\lesssim 10^7$ years or if the NS merger rate is at the higher end of the allowed values, the decrease in median [r-process/Fe] is relatively modest until [Fe/H] $\lesssim -2.5$ in our highest resolution simulation (see Fig.~\ref{fig:models_z2p4}). Moreover, these constraints need only apply at $z \gtrsim 3$ when most of the low-metallicity stars form (see Fig.~\ref{fig:zradius}).

A second uncertainty in our analysis is that we have not implemented NS kicks, which may also contribute to redistributing r-process nuclei relative to heavy elements produced in supernovae. In this respect, our calculations may also \emph{underestimate} the mixing of r-process nuclei. Based on the modest offsets of most short-duration GRBs relative to their host galaxies \citep{Fong2013} we suspect that NS kicks are not likely to be important for the majority of r-process nuclei produced in NS mergers. However, they may be important for enriching low metallicity halo gas. This clearly needs to be explored in detail in future work.

In summary, we have explored a range of values for the NS merger rate per unit stellar mass and the delay time for the first NS mergers to take place, all reasonably consistent with current observational constraints. Our results for the r-process abundance ratios are broadly consistent with available observations. The abundance ratios of r-process elements are not fully converged with the resolution of the simulation, especially at low metallicity. However, variations with resolution are of similar magnitude as variations between different allowed models for the NS merger rate. Considering this as well as the additional uncertainties discussed above, we conclude that NS mergers could well be the source of the majority of the r-process elements in nature.

\section*{Acknowledgements}

We would like to thank Dan Kasen, Patrick Fitzpatrick, and Brian Metzger for useful conversations and the anonymous referee for helpful comments. This work was supported in part by NASA grant NNX10AJ96G, NSF grant AST-1206097, the David and Lucile Packard Foundation, and by a Simons Investigator Award from the Simons Foundation to EQ. DK is supported in part by Hellman Fellowship at UCSD and NSF grant AST-1412153. CAFG is supported by NASA through Einstein Postdoctoral Fellowship Award number PF3-140106 and by NSF through grant number AST-1412836. The simulations here used computational resources granted by the Extreme Science and Engineering Discovery Environment (XSEDE), which is supported by National Science Foundation grant number OCI-1053575; specifically allocations TG-AST120025 (PI Kere\v{s}), TG-AST130039 (PI Hopkins).

\bibliographystyle{mn2e}
\bibliography{rprocess}

\bsp

\label{lastpage}

\end{document}